\documentclass[aip,reprint]{revtex4-1}

\usepackage{amsmath,amsfonts}
\usepackage{bm}
\usepackage{graphicx}
\usepackage{braket}
\usepackage{comment}

\newcommand{\LO}{\mathcal{O}}

\newcommand{\ud}{\mathrm{d}}

\newcommand{\p}{^{\prime}}
\newcommand{\ps}{^{\prime 2}}
\newcommand{\E}[1]{\Braket{#1}}

\begin{document}


\title{Horizontal transport in the bouncing ball system with a sawtooth-shaped table}



\author{Yudai Okishio}
\email[]{aeda1789@chiba-u.jp}
\affiliation{Department of Physics, Chiba University, Yayoi-cho 1-33, Inage-ku, Chiba 263-8522, Japan}
\author{Hiroaki Ito}
\affiliation{Department of Physics, Chiba University, Yayoi-cho 1-33, Inage-ku, Chiba 263-8522, Japan}
\author{Hiroyuki Kitahata}
\email[]{kitahata@chiba-u.jp}
\affiliation{Department of Physics, Chiba University, Yayoi-cho 1-33, Inage-ku, Chiba 263-8522, Japan}

\date{\today}

\begin{abstract}
  The system that consists of a ball bouncing off a sawtooth-shaped table vibrating vertically is considered.
  The horizontal motion in this system is caused by the table shape, and we plotted the mean horizontal velocity as a function of the asymmetry of the table shape.
  The ball is transported in the direction which the gentler slopes face.
  To give a description to the net asymmetric transport,
  we derived a simplified model using assumptions of high bounce and probabilistic collision to the left or right slopes.
  The simplified model exhibits a horizontal transport qualitatively similar to that observed in the original model.
\end{abstract}


\maketitle 

\section{Introduction}

Vibrated granular materials show interesting behavior, such as heaping \cite{Evesque1989}, size separation \cite{Knight1993}, bubbling \cite{Pak1994}, and convection \cite{Ehrichs1995}.
One of the simplest vibrated systems is a bouncing ball system, in which a ball bounces off the table vibrating in a vertical direction under the gravitational field.
This system has been intensively studied in the context of the dynamical systems because the system is simple yet shows rich dynamics, for example, phase locking, sticking, and chaos \cite{Holmes1982, Luck1993, Luo1996, Vogel2011}.
While a horizontally flat shape has been often considered for a table shape in bouncing ball systems,
interesting horizontal dynamics are observed in more realistic systems, in which a dumbbell-like object instead of a ball \cite{Dorbolo2005, Dorbolo2009, Kubo2015}
and a non-flat table instead of a flat table \cite{McBennett2016, Derenyi1998, Farkas1999, Levanon2001, Cai2019, Bae2004, Halev2018} are considered.

Among these systems, a bouncing ball system with a sawtooth-shaped table \cite{Derenyi1998, Farkas1999, Levanon2001, Cai2019, Bae2004, Halev2018} is an important example of granular transport.
This table shape is introduced from the profile of the potential used in the Brownian ratchet \cite{Astumian1994}, which also has been considered as a typical mechanism for horizontal transport from a symmetric energy source.
In multiple-particle systems, the horizontal transport was observed in both positive and negative directions at the same asymmetry on the table shape, depending on the parameters such as the vibration frequency and the height of sawtooth \cite{Farkas1999}.
In single-particle systems, horizontal transport in both directions was reported as well.
When the table shape is composed of vertical walls and left-facing slopes, 
the ball is transported in the left and right directions depending on the restitution coefficient \cite{Bae2004}.
The periodic bounces of a ball on the table whose shape is approximated with low-wavenumber-mode Fourier series were also studied and net horizontal motion in both directions can be achieved depending on the initial condition \cite{Halev2018}.
However, the direction in which even a single ball moves at a certain parameter set and its mechanism have not been clarified yet.

In the present study, we investigate the motion of a single ball bouncing off a sawtooth-shaped table.
We first check the horizontal transport through a numerical calculation.
Then, we analytically derive a simplified model under certain approximations and qualitatively reproduce the horizontal transport using this model. 
We also confirm the validity of the approximations and discuss the mechanism of the dependence of the mean horizontal velocity on the table shape asymmetry.

\section{Model}
We consider a system that consists of an infinitely small ball bouncing off a massive table in the gravitational field. 
We set the gravitational acceleration as $-a_g ~(a_g>0)$.
The table is vibrated periodically in the vertical direction as a function of time $t$. 
Let the horizontal position and velocity of the ball be $x$ and $u$, respectively,
and its vertical position and velocity be $z$ and $v$, respectively.
We set dimensionless forms with a characteristic time $T$ and length $a_gT^2$, where $T$ is the period of the table vibration, as follows:
\begin{align}
  \tilde{t} = t/T,~~
  &\tilde{u} = u/(a_gT),~~ \tilde{x} = x/(a_gT^2), \notag \\
  &\tilde{v} = v/(a_gT),~~ \tilde{z} = z/(a_gT^2).
\end{align}
We omit the tildes in the succeeding descriptions for simplicity.
Assuming that the table is not horizontally flat,
the vertical position of the table $h(t, x)$ is described with a time variation of a baseline of the table $f(t)$ and 
a shape of the table $g(x)$ as follows:
\begin{align}
  h(t, x) = f(t) + g(x).
\end{align}
We use the dot and the prime for time and spatial derivatives like $\dot{f}(t)$ for $\ud f(t)/\ud t$ and $g\p(x)$ for $\ud g(x)/\ud x$, respectively.

Between collisions, the dynamics of the ball is determined by the equation of motion $\ddot{z}=-1$.
We define $t_i$ and $x_i$ as the time and horizontal position at the $i$-th collision, respectively.
We also set the horizontal and vertical reflect velocities at the $i$-th collision to $u_i$ and $v_i$, respectively.
Notably, the vertical position $z_i$ is not included as a state variable since $z_i=h(t_{i}, x_{i})$ holds for each $i$.
With the set of the state variables $t_i, x_i, u_i$ and $v_i$ as an initial condition of the equation of motion, we have the trajectory between the $i$-th and $(i+1)$-th collisions. 
The vertical position of the ball should be the same as the vertical position of the floor at the next collision time.
To find this time, therefore, we solve the following equation
\begin{align}
  -\frac{1}{2}(t-t_i)^2 + v_i(t-t_i) + h(t_i, x_i) = h(t, x_i+u_i(t-t_i)). \label{eq_t}
\end{align}
The $(i+1)$-th collision time $t_{i+1}$ is the minimum value, which is larger than the $i$-th collision time $t_i$ and satisfies Eq. (\ref{eq_t}).

Besides, we derive a relation between the incident and reflect velocities involved in a collision.
The relation in a collision with the floor that has the shape of $g(x)$ only depends on the tilt of the slope $g\p(x)$ at the collision point.
We introduce the tilt angle $\theta$ as $\tan\theta = g^{\prime}(x)$ and then the rotation matrix is defined by
\begin{align}
  R(\theta) = 
  \left(
  \begin{array}{cc}
    \cos\theta & -\sin\theta \\
    \sin\theta & \cos\theta
  \end{array}
  \right).
\end{align}
We assume that a collision can be described with the coefficients of restitution 
and set the tangential and normal coefficients of restitution to unity and $r~(0<r<1)$, respectively.
Considering the rotation of the coordinates, we obtain
\begin{align}
  R(-\theta)
  \left(
  \begin{array}{c}
    u_i \\
    v_i
  -\dot{f}(t_i)
  \end{array}
  \right)
  = 
  \left(
  \begin{array}{cc}
    1 & 0 \\
    0 & -r
  \end{array}
  \right)
  R(-\theta)
  \left(
  \begin{array}{c}
    u_i^* \\
    v_i^*
  -\dot{f}(t_i)
  \end{array}
  \right),
\end{align}
where $u_i^*$ and $v_i^*$ are the $x$ and $z$ components of the incident velocities at the $i$-th collision, respectively.
Using the next incident velocities $u_{i+1}^* = u_i$ and $v_{i+1}^* = v_i - \Delta t_i$, 
where $\Delta t_i = t_{i+1} - t_i$,
we obtain the discrete dynamical system as follows:
\begin{subequations}
\label{ds}
\begin{align}
  t_{i+1} &= \mathrm{min} {\huge \mbox{$\{$}} t\in(t_i, \infty) ~{\huge \mbox{$|$}}~ \notag \\ 
  & -\cfrac{1}{2}(t-t_i)^2 + v_i(t-t_i) + h(t_i, x_i)
  = h(t, x_{i+1}) {\huge \mbox{$\}$}}, \\
  x_{i+1} &= x_i + u_i\Delta t_i, \\
  u_{i+1} &= \cfrac{1-rg\ps_{i+1}}{1+g\ps_{i+1}}u_i + 
  (1+r)\cfrac{g\p_{i+1}}{1+g\ps_{i+1}} (v_i-\Delta t_i - \dot{f}_{i+1}), \\
  v_{i+1} &= (1+r)\cfrac{1}{1+g\ps_{i+1}}(g\p_{i+1} u_i + \dot{f}_{i+1})  \notag \\
  & ~~~~~ + \cfrac{g\ps_{i+1} - r}{1+g\ps_{i+1}}(v_i-\Delta t_i),
\end{align} 
\end{subequations}
where $\dot{f}_{i+1} = \dot{f}(t_{i+1})$ and $g\p_{i+1} = g\p(x_{i+1})$.

\begin{figure}[tbp]
  \begin{center}
    \includegraphics{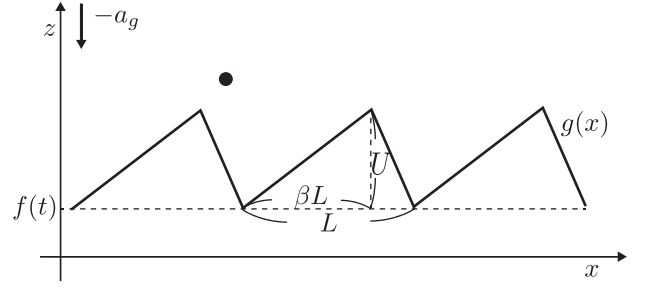}
    \caption{Schematic illustration of the present system.}
    \label{fig:1}
  \end{center}
\end{figure}

We adopt a sinusoidal function with an amplitude $\alpha$ for $f(t)$ as in a typical one-dimensional bouncing ball model\cite{Holmes1982}.
As for the shape $g(x)$, we adopt a sawtooth-shaped function, which is defined by three parameters, 
an asymmetry $\beta$, a width of each tooth $L$, and a height of each tooth $U$ as shown in Fig. \ref{fig:1}.
These functions are explicitly denoted by
\begin{align}
  f(t) &= \alpha \sin 2\pi t, \\
  g(x) &= \left\{
    \begin{array}{ll}
      \gamma_l\left(x - L\left\lfloor\frac{x}{L}\right\rfloor\right), 
      & \frac{x}{L} - \lfloor\frac{x}{L}\rfloor<\beta, \\
      \gamma_r\left(x - L\left\lfloor\frac{x}{L}\right\rfloor-L \right), 
      & \frac{x}{L} - \lfloor\frac{x}{L}\rfloor \ge \beta,
    \end{array}
  \right.
\end{align}
where $\lfloor \cdot \rfloor$ is the floor function, and $\gamma_l=U/(\beta L)$ and $\gamma_r=-U/\{(1-\beta)L\}$.

\section{Results}
To quantitatively evaluate the horizontal transport, we introduce the mean horizontal velocity (MHV)
\begin{align}
  \bar{u} = \E{\frac{\sum_i u_i\Delta t_i}{\sum_i \Delta t_i}}, \label{MHV}
\end{align}
where $\E{\cdot}$ represents the average of the values after sufficiently long time with respect to initial conditions. 
Since we cannot analytically solve the dynamical system in Eq. (\ref{ds}), we numerically calculated MHV.
Besides, we approximately simplified the original model and analytically estimated MHV with the simplified model.

\subsection{Numerical result}
We performed numerical calculation to obtain MHV as a function of $\beta$.
We simulated the dynamics until 10000 collisions for each run and discarded the data of initial 9000 collisions.
The solver uses the bisection method for finding a collision time.
We regarded the solution with serial small flight durations ($\Delta t_i < 10^{-8}$) as a sticking solution \cite{Vogel2011} and did not include the sticking solutions in the calculation of MHV.
The dynamics were computed from 1000 initial horizontal positions $x_0$ evenly distributed between $0$ and $L$, where the sticking solutions were observed from some of the initial positions.
For the other initial conditions, $z_0=1$ and $u_0=v_0=0$ were always adopted. 
Note that we did not determined $t_0$ but determined $z_0$ and started the computation from $t=0$ so that the vertical positions of the ball and the table were not the same at an initial condition.
We calculated collisions for $100$ points of the asymmetry $\beta$ evenly distributed between $0$ and $1$.

\begin{figure*}[tbp]
  \begin{center}
    \includegraphics{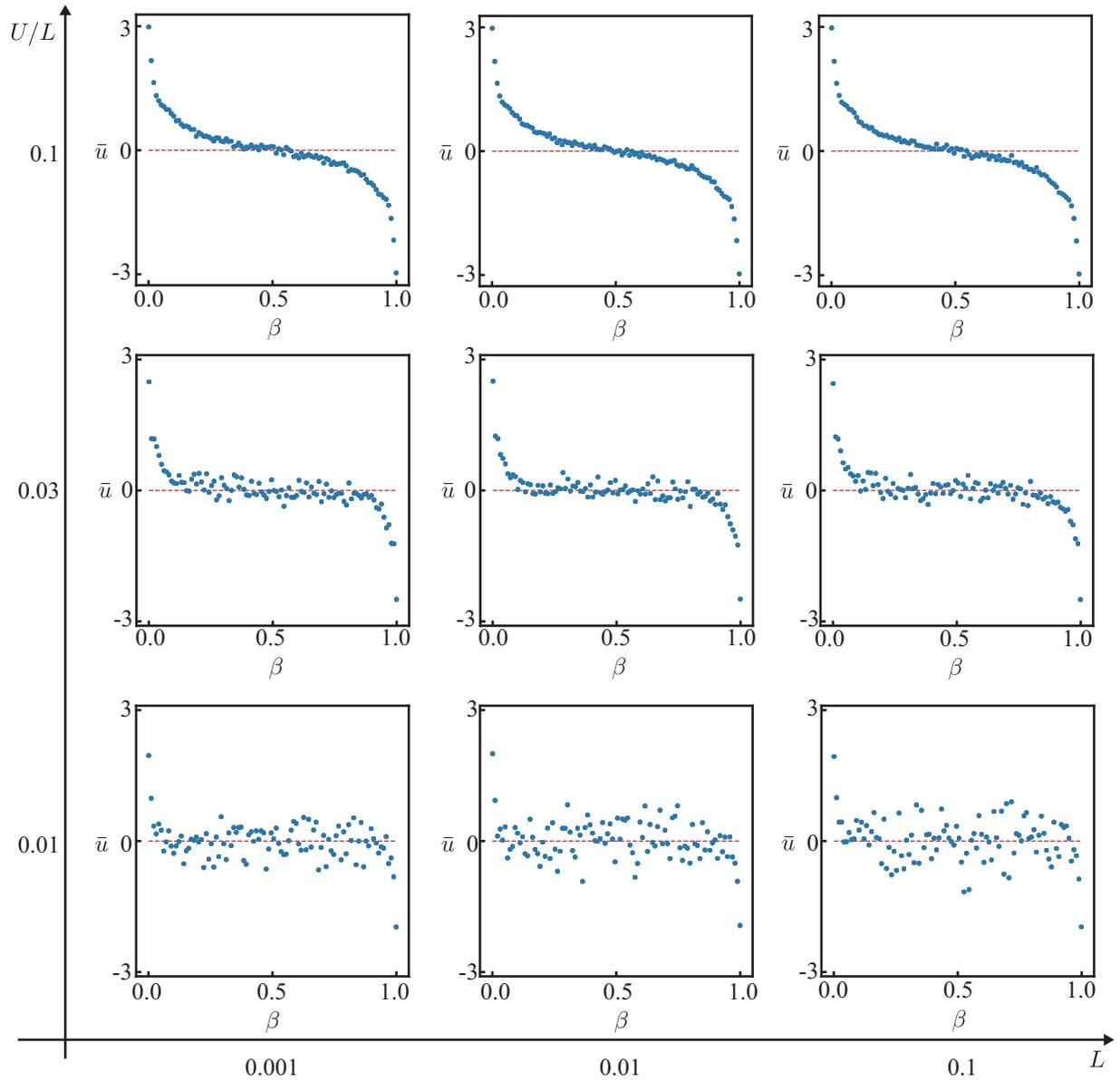}
    \caption{MHV $\bar{u}$ against the asymmetry of the table shape $\beta$ obtained from the numerical calculation using the originial model in Eq. (\ref{ds}) for the amplitude of vibration $\alpha=1$ and the coefficient of restitution $r=0.8$.
    The red dashed lines represent $\bar{u}=0$.
    }
    \label{fig:2}
  \end{center}
\end{figure*}

We plotted MHV as a function of $\beta$ for various $U/L$ and $U$ under the fixed parameters $\alpha = 1$ and $r = 0.8$ in Fig. \ref{fig:2}.
It shows that MHV clearly depends on $\beta$ in the wide range of $L$ when $U/L$ is sufficiently large.
The ball is transported in the positive direction when $\beta<0.5$ and in the negative direction when $\beta>0.5$.
The dependence on asymmetry $\beta$ is nonlinear.

\subsection{Derivation of simplified model}
We simplify the original model in Eq. (\ref{ds}) to derive MHV analytically under a certain approximation.
The difficulties in analyzing the original model are that the equation is implicit for $t_i$ and that the number of state variables is four.
Eliminating some of state variables, particularly $t_i$, helps us to handle this model analytically.
Thus, we make the following three assumptions: 
(i) high bounce, 
(ii) uniform horizontal distribution of incident positions over a sawtooth,
and (iii) constant incident velocities.
As explained below, assumptions (i) and (ii) eliminate $t_i$ and $x_i$, respectively, from the original model in Eq. (\ref{ds}).
Furthermore, assumption (iii) also eliminates the state variable $v_i$ and is needed to be consistent with assumptions (i) and (ii).
As a consequence, we obtain a probabilistically switching dynamical system for a single state variable $u_i$.

First, we eliminate $t_i$ from the original model in Eq. (\ref{ds}).
We consider that each bounce height is sufficiently larger than the amplitudes of the table shape and the vibration.
Then Eq. $(\ref{eq_t})$ is approximated by
\begin{align}
  -\frac{1}{2}(t-t_i)^2 + v_i (t-t_i) = 0.
\end{align}
This equation yields
\begin{align}
  \Delta t_i = 2v_i. \label{hb}
\end{align}
In addition, the incident velocities are described by 
\begin{align}
  \begin{array}{l}
    u_i^* = u_i, \\
    v_i^* = v_i - \Delta t_i = -v_i.
  \end{array} \label{hba_uv}
\end{align}
The high bounce approximation in Eq. (\ref{hb}) enables us to avoid considering the implicit equation for $t_i$ in Eq. (6). 

\begin{figure}[tbp]
  \begin{center}
    \includegraphics{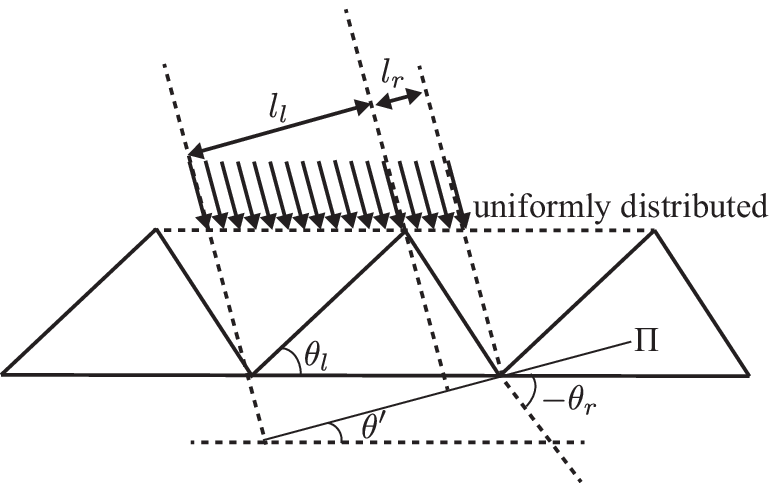}
    \caption{Schematic image of the projection of slopes onto plane $\Pi$.}
    \label{fig:3}
  \end{center}
\end{figure}

Second, we ignore the horizontal position $x_i$ in Eq. (\ref{ds}) under the high bounce approximation.
Since $x_i$ appears only through $g\p(x_i)$, which takes only one out of two values $\gamma_l$ and $\gamma_r$, we do not need the actual value of $x_i$.
Here, $\gamma_l$ and $\gamma_r$ are the spatial derivatives of the left and right slopes, respectively, as in Eq. (8).
We consider the dynamics in a probabilistic perspective and assume that $x_i$ is uniformly distributed over a sawtooth.
Let plane $\Pi$ be perpendicular to a velocity vector as shown in Fig. \ref{fig:3}.
Probabilities of a collision with the left and right slopes are calculated from the ratio of the projections of the slopes onto plane $\Pi$.
Let the tilt angles of the left and right slopes of the table be $\theta_l$ and $\theta_r$, respectively, where
\begin{align}
  \tan\theta_\nu = \gamma_\nu, ~ -\frac{\pi}{2}\le \theta_\nu <\frac{\pi}{2}.
\end{align}
Here, $\nu$ denotes $l$ or $r$.
Besides, let the tilt angle of plane $\Pi$ be $\theta\p$, which is calculated as
\begin{align}
  \tan\theta\p = \frac{u_i^*}{-v_i^*} = \frac{u_i}{v_i}, ~ -\frac{\pi}{2}\le \theta\p <\frac{\pi}{2}.
\end{align}
These tilt angles $\theta_l, \theta_r$, and $\theta\p$ are measured counterclockwise from the $x$-axis.
Projections of the left and right slopes onto plane $\Pi$ are given as
\begin{align}
  l_l &= \sqrt{(\beta L)^2 + U^2}\cos(\theta_l-\theta\p) \notag \\
  &= \frac{1}{\sqrt{u_i^{2}+v_i^{2}}}(\beta Lv_i + Uu_i), \\
  l_r &= \sqrt{\{(1-\beta) L\}^2 + U^2}\cos(-\theta_r+\theta\p) \notag \\
  &= \frac{1}{\sqrt{u_i^{2}+v_i^{2}}}((1-\beta) Lv_i - Uu_i).
\end{align}
Here, either $l_l$ or $l_r$ can be negative.
In such a case, the probability of a collision with the corresponding side of slopes is set to be zero.
Therefore, the probability that the ball collides with the left slopes $p_l$ is summarized as
\begin{align}
  p_l = \left\{
  \begin{array}{ll}
    0, & l_l < 0, \\
    1, & l_r < 0, \\
    \frac{l_l}{l_l+l_r} = \beta + A\frac{u_i}{v_i}, & \mathrm{otherwise},
  \end{array} \right. \label{p}
\end{align}
where $A\equiv U/L$ is the aspect ratio of the table shape.

Assumption (i) does not hold when $v_i$ is small.
A collision cannot occur when $v_i$ is negative under assumption (ii).
Therefore we make assumption (iii) that $v_i$ is a positive constant at a sufficiently large value $v$, which helps assumptions (i) and (ii) always be fulfilled and further reduces the variable $v_i$.

Hence, the model is simplified to a one-dimensional probabilistic dynamical system
\begin{align}
  u_{i+1} = \left\{
  \begin{array}{ll}
    a_lu_i + b_l, & \mathrm{prob.} = p_l, \\
    a_ru_i + b_r, & \mathrm{prob.} = 1-p_l, \label{1d_model}
  \end{array}
  \right.
\end{align}
where the coefficients are described as
\begin{align}
  a_\nu &= \frac{1-r\gamma^2_\nu}{1+\gamma^2_\nu}, \label{a_nu} \\
  b_\nu &= -(1+r)\frac{\gamma_\nu}{1+\gamma^2_\nu} v,
\end{align}
where $\nu = l, r$.
We ignore the term of $\dot{f}_{i+1}$ in the derivation of Eq. (\ref{1d_model}) because of the high bounce approximation, which results in the complete elimination of $t_i$ from the model.

\begin{figure}[tbp]
  \begin{center}
    \includegraphics[width=\linewidth]{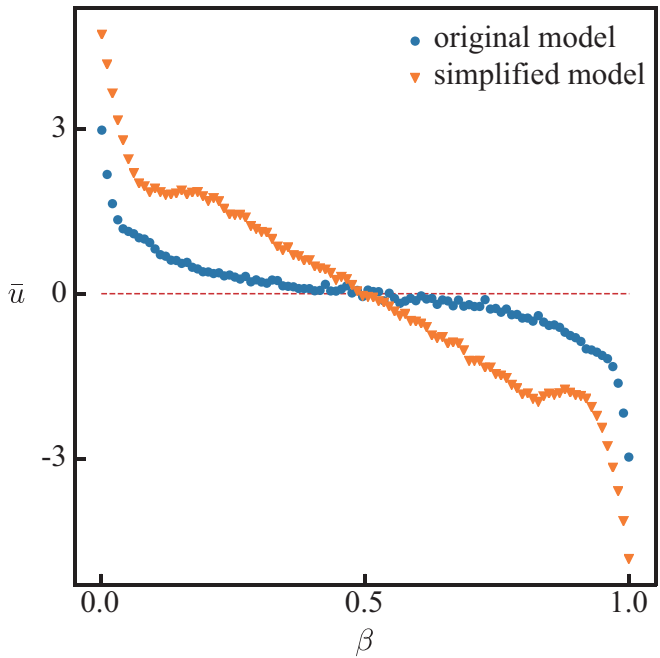}
    \caption{
      MHV numerically calculated using the simplified model discribed by probabilistic one-dimensional model in Eq. (\ref{1d_model}) (orange) and using the original model (blue) in Eq. (\ref{ds}).
      The amplitude of vibration $\alpha=1$ and the coefficient of restitution $r=0.8$, which are the same as those in Fig. \ref{fig:2}.
      In the simplified model, we adopt $L=0.1$ and $A=0.1$ and assume $v=14.7$, which is the mean vertical velocity obtained from the numerical calculation using the original model.
      The red dashed line represents $\bar{u}=0$.
    }
    \label{fig:4}
  \end{center}
\end{figure}

Since the expected value of $u_i$ cannot be calculated analytically, we first perfomed the numerical calculation.
Figure \ref{fig:4} shows that the dependence of MHV on $\beta$ in the simplified model. 
It is similar to that in the original model despite ignoring three out of the four variables.
We adopt $v=14.7$, which is the mean value of vertical incident velocities obtained from the numerical calculation using the original dynamical system.

\subsection{Derivation of MHV in the simplified model}
We assume that the invariant measure $P(u)$ exists in the simplified model in Eq. (\ref{1d_model}).
If $u$ is rarely in the region where $p_l$ is equal to $0$ or $1$, then we obtain an equation on the probability flow as follows:
\begin{align}
  P(u)\ud u &= \left(\beta+\frac{A}{v}u\p_l\right)P(u\p_l)\ud u\p_l \notag \\
  &~~~ +\left\{(1-\beta)-\frac{A}{v}u\p_r\right\}P(u\p_r)\ud u\p_r, \label{prob_dist_eq}
\end{align}
where $u = a_lu\p_l+b_l = a_ru\p_r+b_r$.
Instead of directly finding the function $P(u)$ from the equation, we assume that $P$ follows the normal distribution.
Due to this, we solve the equations for mean $\mu$ and variance $\sigma^2$.
Integrating the product of Eq. (\ref{prob_dist_eq}) and $u$ or $u^2$ yields the complicated cubic equation for $\mu$ in Eq. (\ref{long_eq}) in the Appendix.
We simplify the equation with the perturbation method around $\beta=1/2$ and find
\begin{align}
  \mu &= \left(\frac{b_1+2b_0}{1-a_0-2(A/v)b_0} \right. \notag \\ &~~~~~ \left. +\frac{A}{v}\frac{2a_1b_0^2}{(a_0^2+4(A/v)a_0b_0-1)
(a_0+2(A/v)b_0-1)}\right)\delta \notag \\ &~~~~~ + \LO(\delta^2), \label{approx_mu}
\end{align}
where $\delta\equiv\beta-1/2$.
Here $a_0, a_1, b_0$, and $b_1$ are the expansion coefficients of $a_l$ and $b_l$ with respect to $\delta$ as
\begin{align}
  a_l &= a_0 + a_1\delta + \LO(\delta^2), \\
  b_l &= b_0 + b_1\delta + \LO(\delta^2), 
\end{align}
which are explicitly described as
\begin{subequations}
\label{coef}
\begin{align}
    a_0 &= \cfrac{1-4rA^2}{1+4A^2}, \\
    a_1 &= (1+r)\cfrac{16A^2}{(1+4A^2)^2}, \\
    b_0 &= -(1+r)v\cfrac{2A}{1+4A^2}, \\
    b_1 &= (1+r)v\cfrac{4A(1-4A^2)}{(1+4A^2)^2}.
\end{align}
\end{subequations}
The detailed derivation is included in the Appendix.
Figure \ref{fig:5} shows that the approximate analytical solution in Eq. (\ref{approx_mu}) matches the numerical result when $\beta$ is nearly $1/2$.
The approximate solution is close to the numerical result especially for $0.1<\beta<0.9$ as shown in Fig. 5.
This could be understood as follows:
When $\beta$ or $1-\beta$ is less than $A\sqrt{r}\simeq 0.0894$, either $a_l$ or $a_r$ is negative from Eq. (\ref{a_nu}).
In this case, $u_i$ more frequently takes the value in the region with $p_l=0$ or $p_l=1$ compared to the case that both $a_l$ and $a_r$ are positive.
Equation (\ref{prob_dist_eq}) is, therefore, no longer good approximation for $\beta<0.1$ or $\beta>0.9$.

\begin{figure}[tbp]
  \begin{center}
    \includegraphics[width=\linewidth]{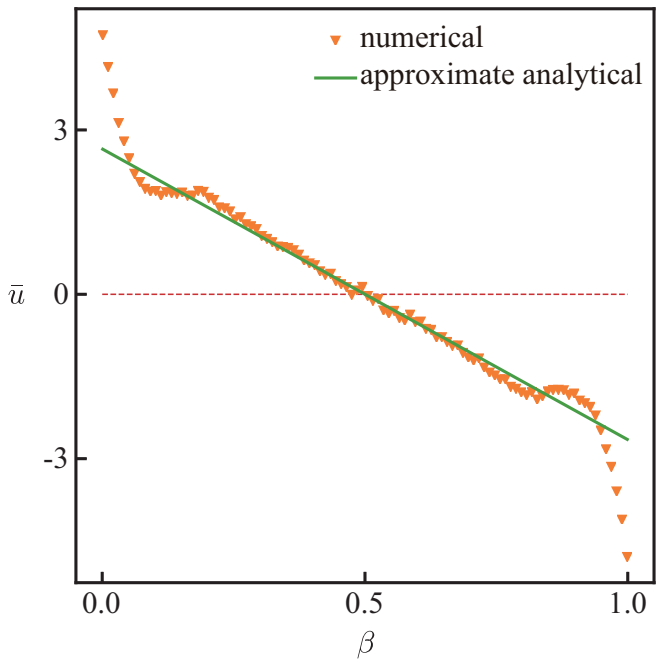}
    \caption{
    MHV from the numerical (orange) and approximate analytical (green) calculations
    in the simpilfied model. 
    The parameters are $\alpha = 1$, $r=0.8$, and $A=0.1$. Note that $L$ is not included in the simplified model.
    We assume $v=14.7$ in both the calculations. The red dashed line represents $\bar{u}=0$.}
    \label{fig:5}
  \end{center}
\end{figure}

\section{Discussion}
In the numerical calculation using the original model, we found that the ball tends to be transported in the direction that the gentler slopes face.
The gentler slope is more difficult to accelerate the ball in the horizontal direction but easier to collide due to the longer length, and the present result suggests that the latter effect exceeds.
We also confirmed that MHV obtained from the simplified model has a similar dependence on the table shape asymmetry $\beta$ to that from the original model. 
The simplest model which we can consider is the one in which the probabilities of collision with the left and right slopes are replaced by $\beta$ and $1-\beta$ in Eq. (\ref{1d_model}), respectively.
Even this simplest model predicts the sign of the horizontal transport calculated with the original model.
However, the value of MHV obtained from the simplest model is one order greater than that from the original model.
The dependence of the collision probability on the direction of the incident velocity vector in our model is critical to deceleration.
It is indicated that the ball is not overaccelerated in the horizontal directon due to ease to collide with the opposite slopes also in the original model.

On the other hand, we did not observe reversal of MHV by varying the parameters as reported in the previous studies \cite{Farkas1999, Bae2004}.
This may be because we adopted the parameter sets at which sticking solutions do not frequently appear.
In such parameter regions, bounces are expected to be high and the mechanism precedingly discussed can be applied.
To the contrary, in the parameter region in which sticking solutions frequently appear, the reversal of MHV may be observed.
The reason of this may be that the reverse transport would be caused by the set of two bounces; a bounce off a gentler slope and the following bounce off the adjacent steeper slope.
In such sequential bounces, the steeper slopes act as a ``wall". 
This sort of sequential bounces is likely to occur when the bounces are low, which is not the case for the parameter sets used in this paper.

Hereinafter, we examine the validity of the three assumptions, 
(i) high bounce, 
(ii) uniform distribution of horizontal position over a tooth,
and (iii) constant incident velocities.
Although these assumptions are used in the derivation of the simplified model without justification,
we will confirm that the assumptions match with the data from the numerical calculation using the original model.
We only plotted the data after sufficiently long time as in the results section.

\begin{figure}[tbp]
  \begin{center}
    \includegraphics[width=\linewidth]{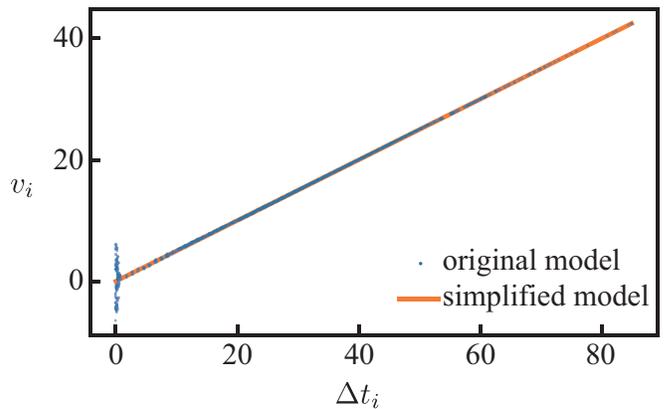}
    \caption{Scatter plot of $v_i$ vs. $\Delta t_i$ by the numerical calculation using the original model in Eq. (\ref{ds}) (blue) and using the simplified model (orange).
    The parameters are $\alpha = 1$, $r=0.8$, $L=0.1$, $A=0.1$, and $\beta=0.404$.
    }
    \label{fig:6}
  \end{center}
\end{figure}

First, we plotted the reflect velocities and flight durations in Fig. \ref{fig:6}.
The plot shows that Eq. (\ref{hb}) holds when $\Delta t$ is large.
Although there is a number of flights with a small duration including sticking solutions, the contribution to MHV from these bounces is relatively small.
Therefore overall this assumption is valid.

\begin{figure}[tbp]
  \begin{center}
    \includegraphics[width=\linewidth]{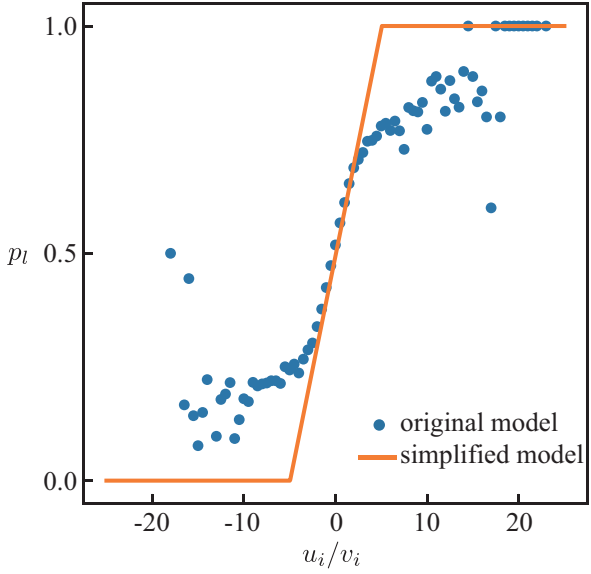}
    \caption{Fraction of the collision with the left slopes in the original model (blue).
    Each of the points is fraction calculated with the range of $0.5$ for $u_i/v_i$.
    We only plotted the data that fulfill $\Delta t_i>3$ corresponding to high bounces for the results with the original model.
    We also plotted the probability used in the simplified model in Eq. (\ref{p}) (orange).
    The parameters are $\alpha = 1$, $r=0.8$, $L=0.1$, $A=0.1$, and $\beta=0.404$.
    }
    \label{fig:7}
  \end{center}
\end{figure}

Second, assumption (ii) is essential to determine the next horizontal velocity in contrast with the assumptions (i) and (iii).
We plotted the fraction of collisions with the left slopes as a function of $u_i/v_i$ in Fig. \ref{fig:7}.
The plot shows that assumption (ii) in Eq. (\ref{p}) is valid under the high bounce approximation.

Finally, we supposed that $v_i$ is constant in assumption (iii) for consistency with the preceding two assumptions.
Due to this constant $v$, the flight duration $\Delta t_i$ is also constant, i.e., all the bounces have the same height.
In the original model, if the bounce has a large horizontal velocity, the bounce should not be high and should immediately collide to the opposite slopes, which decelerate the ball in the horizontal direction.
The difficulty in deceleration in the simplified model may be attributed to the difference from the results using the original model.

\section{Conclusion}
We investigated the horizontal transport of a ball bouncing off a sawtooth-shaped table vibrating vertically by both numerical and approximate analytical calculations.
In the case that the aspect ratio of the table shape $A$ is sufficiently large, MHV (mean horizontal velocity) clearly depends on asymmetry $\beta$ of the table shape.
The ball is transported in the horizontal direction that the gentler slopes face.
We reproduced such dependence using the probabilistic model approximately derived from the relation between the direction of velocity and tilt of the slope.
Our analysis suggests that the effect of more frequent collisions with the gentler slopes exceeds that of the lower magnitude of the horizontal acceleration by the gentler slopes, though MHV has a moderated value since the probability of a collision with the steeper slopes become higher for the larger horizontal velocity.
The direction of the transport would change for the parameter set at which sticking solutions more frequently appear.
Considering the constraint condition for the sticking solutions would enable us to analyze motion of the ball after sticking and to further understand the mechanism of the transport reversal.

\section{Acknowledgments}
This work was supported by JST SPRING, Grant Number JPMJSP2109 (YO).
This work was also supported by JSPS KAKENHI Grant Nos. JP19H00749, JP21K13891 (HI), JP20H02712, JP21H00996, and JP21H01004 (HK), 
and also the Cooperative Research Program of ``Network Joint Research Center for Materials and Devices: Dynamic Alliance for Open Innovation Bridging Human, Environment and Materials'' No. 20214004 (HK).
This work was also supported by JSPS and MESS Japan-Slovenia Research Cooperative Program Grant No. JPJSBP120215001 (HI), and
JSPS and PAN under the Japan-Poland Research Cooperative Program No. JPJSBP120204602 (HK).

\appendix
\begin{widetext}
\section{}

In the Appendix, we derive Eq. (\ref{approx_mu}) in detail.
By multiplying Eq. (\ref{prob_dist_eq}) with $u$,
\begin{align}
  uP(u)\ud u &= \left(\beta u+\frac{A}{v}uu\p_l\right)P(u\p_l)\ud u\p_l 
  + \left\{(1-\beta)u-\frac{A}{v}uu\p_r\right\}P(u\p_r)\ud u\p_r \notag \\
  &= \left(\beta (a_lu\p_l+b_l)+\frac{A}{v}(a_lu\p_l+b_l)u\p_l\right)P(u\p_l)\ud u\p_l 
  + \left\{(1-\beta)(a_ru\p_r+b_r)-\frac{A}{v}(a_ru\p_r+b_r)u\p_r\right\}P(u\p_r)\ud u\p_r.
\end{align}
Integrating and simplifying this equation, we have
\begin{align}
  \frac{A}{v}(a_l-a_r)(\mu^2+\sigma^2)
  + \left\{a_l\beta+a_r(1-\beta)+\frac{A}{v}(b_l-b_r)-1\right\}\mu
  + b_l \beta + b_r(1-\beta) = 0. \label{moment1}
\end{align}
By multiplying Eq. (\ref{prob_dist_eq}) by $u^2$ and simplifying it,
\begin{align}
  &\frac{A}{v}(a_l^2-a_r^2)(\mu^2+3\sigma^2)\mu \notag
  \\&~~~~~ + \left\{a_l^2\beta+a_r^2(1-\beta)+2\frac{A}{v}(a_lb_l-a_rb_r)-1 \right\}(\mu^2+\sigma^2) \notag
  \\&~~~~~ + \left\{2a_lb_l\beta+2a_rb_r(1-\beta)+\frac{A}{v}(b_l^2-b_r^2) \right\} \mu \notag
  \\&~~~~~ + b_l^2\beta+b_r^2(1-\beta) = 0,
\end{align}
where we use the assumption that $u$ follows the normal distribution with the mean $\mu$ and the variance $\sigma^2$.
Substituting $\sigma^2$ derived from Eq. (\ref{moment1}) to this, we obtain
\begin{align}
  & -2\frac{A}{v}(a_l^2-a_r^2)\mu^3 \notag
  \\&~~~~~ -3(a_l+a_r)\left\{a_l\beta+a_r(1-\beta)+\frac{A}{v}(b_l-b_r)-1\right\}\mu^2 \notag
  \\&~~~~~ + \left[2a_lb_l\beta+2a_rb_r(1-\beta)+\frac{A}{v}(b_l^2-b_r^2)-3(a_l+a_r)\{b_l\beta+b_r(1-\beta)\}\right. \notag
  \\&~~~~~ -\frac{v}{A}\frac{1}{a_l-a_r} \notag
  \left. \left\{a_l^2\beta+a_r^2(1-\beta)+2\frac{A}{v}(a_lb_l-a_rb_r)-1 \right\}
  \left\{a_l\beta+a_r(1-\beta)+\frac{A}{v}(b_l-b_r)-1\right\}
  \right] \mu
  \\&~~~~~ + b_l^2\beta+b_r^2(1-\beta)
  -\frac{v}{A}\frac{1}{a_l-a_r}
  \left\{a_l^2\beta+a_r^2(1-\beta)+2\frac{A}{v}(a_lb_l-a_rb_r)-1 \right\}
  \left\{b_l \beta + b_r(1-\beta)\right\}
  = 0. \label{long_eq}
\end{align}
We perform the Taylor expansion with respect to  $\delta=\beta-1/2$ for this equation.
The preceding equation (\ref{long_eq}) contains $\beta$ implicitly through $a_l, a_r, b_l$ and $b_r$ as well as explicitly.
We expand $a_l$ and $b_l$ with substituting the expansion coefficients into $a_0, a_1, b_0$ and $b_1$ as 
\begin{align}
  a_l &= a_0 + a_1\delta + \LO(\delta^2), \\
  b_l &= b_0 + b_1\delta + \LO(\delta^2).
\end{align}
We can also expand $a_r$ and $b_r$ using the same expansion coefficients for $a_l$ and $b_l$ as
\begin{align}
  a_r &= a_0 - a_1\delta + \LO(\delta^2), \\
  b_r &= -b_0 + b_1\delta+ \LO(\delta^2).
\end{align}
Here we use Eq. (\ref{coef}).
Now we solve Eq. (\ref{long_eq}) with the perturbation method upto the first order of $\delta$.
We expand $\mu$ as
\begin{align}
  \mu = \mu_0 + \mu_1 \delta + \LO(\delta^2).
\end{align}
Since Eq. (\ref{long_eq}) contains the division by $a_l-a_r=a_1\delta$, the equation of the zero-th order of $\delta$ should be included in the following expression:
\begin{align}
  & \left\{a_l^2\beta+a_r^2(1-\beta)+2\frac{A}{v}(a_lb_l-a_rb_r)-1 \right\}
  \left\{a_l\beta+a_r(1-\beta)+\frac{A}{v}(b_l-b_r)-1\right\}
  \mu_0 \notag \\
  &~~~ -\left\{a_l^2\beta+a_r^2(1-\beta)+2\frac{A}{v}(a_lb_l-a_rb_r)-1 \right\}
  \left\{b_l \beta + b_r(1-\beta)\right\}.
\end{align}
The constant term in the zero-th order of $\delta$ is zero because
\begin{align}
  b_l \beta + b_r(1-\beta) &= (b_1 + 2b_0)\delta.
\end{align}
Then, we have $\mu_0=0$ because the coefficient of $\mu_0$ has a non-zero value.

We consider the first-order equation with respect to $\delta$. By $\mu_0 = 0$, we ignore the second- or higher-order terms of $\mu$.
\begin{align}
  & \left[(a_l-a_r)\left(2a_lb_l\beta+2a_rb_r(1-\beta)+\frac{A}{v}(b_l^2-b_r^2)-3(a_l+a_r)\{b_l\beta+b_r(1-\beta)\}\right)\right. \notag
  \\&~~~~~ -\frac{v}{A}
  \left. \left\{a_l^2\beta+a_r^2(1-\beta)+2\frac{A}{v}(a_lb_l-a_rb_r)-1 \right\}
  \left\{a_l\beta+a_r(1-\beta)+\frac{A}{v}(b_l-b_r)-1\right\}
  \right] \mu \notag
  \\&~~~~~ + (a_l-a_r)(b_l^2\beta+b_r^2(1-\beta))
  -\frac{v}{A}
  \left\{a_l^2\beta+a_r^2(1-\beta)+2\frac{A}{v}(a_lb_l-a_rb_r)-1 \right\}
  \left\{b_l \beta + b_r(1-\beta)\right\}
  = 0.
\end{align}
Now, considering $\mu = \mu_1\delta$, the first-order terms of $\delta$ should be in the following expression:
\begin{align}
  & -\frac{v}{A}
  \left\{a_l^2\beta+a_r^2(1-\beta)+2\frac{A}{v}(a_lb_l-a_rb_r)-1 \right\}
  \left\{a_l\beta+a_r(1-\beta)+\frac{A}{v}(b_l-b_r)-1\right\}
  \mu_1\delta \notag
  \\&~~~~~ + (a_l-a_r)(b_l^2\beta+b_r^2(1-\beta))
  -\frac{v}{A}
  \left\{a_l^2\beta+a_r^2(1-\beta)+2\frac{A}{v}(a_lb_l-a_rb_r)-1 \right\}
  \left\{b_l \beta + b_r(1-\beta)\right\}
  = 0.
\end{align}
By expanding $a_l, a_r, b_l$ and $b_r$ with respect to $\delta$, we obtain
\begin{align}
  & -\frac{v}{A}
  \left\{\left(a_0^2+4\frac{A}{v}a_0b_0 - 1\right)\left(a+2\frac{A}{v}b_0-1\right) + \LO(\Delta\beta^2)\right\}
  \mu_1\delta \notag \\
  &~~~~~ + 2a_1b_0^2\delta + \LO(\delta^2)
  -\frac{v}{A}
  \left\{2\left(a_0^2+4\frac{A}{v}a_0b_0 - 1\right)b_0\delta + \LO(\delta^2)\right\}
  = 0.
\end{align}
Finally we have
\begin{align}
  \mu_1 &= \frac{b_1+2b_0}{1-a_0-2(A/v)b_0} + \frac{A}{v}\frac{2a_1b_0^2}{(a_0^2+4(A/v)a_0b_0-1)
(a_0+2(A/v)b_0-1)}.
\end{align}

\end{widetext}


%
%

%


\bibliographystyle{apsrev4-1.bst}
\bibliography{ref}

\begin{thebibliography}{19}%
\makeatletter
\providecommand \@ifxundefined [1]{%
 \@ifx{#1\undefined}
}%
\providecommand \@ifnum [1]{%
 \ifnum #1\expandafter \@firstoftwo
 \else \expandafter \@secondoftwo
 \fi
}%
\providecommand \@ifx [1]{%
 \ifx #1\expandafter \@firstoftwo
 \else \expandafter \@secondoftwo
 \fi
}%
\providecommand \natexlab [1]{#1}%
\providecommand \enquote  [1]{``#1''}%
\providecommand \bibnamefont  [1]{#1}%
\providecommand \bibfnamefont [1]{#1}%
\providecommand \citenamefont [1]{#1}%
\providecommand \href@noop [0]{\@secondoftwo}%
\providecommand \href [0]{\begingroup \@sanitize@url \@href}%
\providecommand \@href[1]{\@@startlink{#1}\@@href}%
\providecommand \@@href[1]{\endgroup#1\@@endlink}%
\providecommand \@sanitize@url [0]{\catcode `\\12\catcode `\$12\catcode
  `\&12\catcode `\#12\catcode `\^12\catcode `\_12\catcode `\%12\relax}%
\providecommand \@@startlink[1]{}%
\providecommand \@@endlink[0]{}%
\providecommand \url  [0]{\begingroup\@sanitize@url \@url }%
\providecommand \@url [1]{\endgroup\@href {#1}{\urlprefix }}%
\providecommand \urlprefix  [0]{URL }%
\providecommand \Eprint [0]{\href }%
\providecommand \doibase [0]{http://dx.doi.org/}%
\providecommand \selectlanguage [0]{\@gobble}%
\providecommand \bibinfo  [0]{\@secondoftwo}%
\providecommand \bibfield  [0]{\@secondoftwo}%
\providecommand \translation [1]{[#1]}%
\providecommand \BibitemOpen [0]{}%
\providecommand \bibitemStop [0]{}%
\providecommand \bibitemNoStop [0]{.\EOS\space}%
\providecommand \EOS [0]{\spacefactor3000\relax}%
\providecommand \BibitemShut  [1]{\csname bibitem#1\endcsname}%
\let\auto@bib@innerbib\@empty
\bibitem [{\citenamefont {Evesque}\ and\ \citenamefont
  {Rajchenbach}(1989)}]{Evesque1989}%
  \BibitemOpen
  \bibfield  {author} {\bibinfo {author} {\bibfnamefont {P.}~\bibnamefont
  {Evesque}}\ and\ \bibinfo {author} {\bibfnamefont {J.}~\bibnamefont
  {Rajchenbach}},\ }\href@noop {} {\bibfield  {journal} {\bibinfo  {journal}
  {Phys. Rev. Lett.}\ }\textbf {\bibinfo {volume} {62}},\ \bibinfo {pages} {44}
  (\bibinfo {year} {1989})}\BibitemShut {NoStop}%
\bibitem [{\citenamefont {Knight}\ \emph {et~al.}(1993)\citenamefont {Knight},
  \citenamefont {Jaeger},\ and\ \citenamefont {Nagel}}]{Knight1993}%
  \BibitemOpen
  \bibfield  {author} {\bibinfo {author} {\bibfnamefont {J.~B.}\ \bibnamefont
  {Knight}}, \bibinfo {author} {\bibfnamefont {H.~M.}\ \bibnamefont {Jaeger}},
  \ and\ \bibinfo {author} {\bibfnamefont {S.~R.}\ \bibnamefont {Nagel}},\
  }\href@noop {} {\bibfield  {journal} {\bibinfo  {journal} {Phys. Rev. Lett.}\
  }\textbf {\bibinfo {volume} {70}},\ \bibinfo {pages} {3728} (\bibinfo {year}
  {1993})}\BibitemShut {NoStop}%
\bibitem [{\citenamefont {Pak}\ and\ \citenamefont
  {Behringer}(1994)}]{Pak1994}%
  \BibitemOpen
  \bibfield  {author} {\bibinfo {author} {\bibfnamefont {H.~K.}\ \bibnamefont
  {Pak}}\ and\ \bibinfo {author} {\bibfnamefont {P.~R.}\ \bibnamefont
  {Behringer}},\ }\href@noop {} {\bibfield  {journal} {\bibinfo  {journal}
  {Nature}\ }\textbf {\bibinfo {volume} {371}},\ \bibinfo {pages} {231}
  (\bibinfo {year} {1994})}\BibitemShut {NoStop}%
\bibitem [{\citenamefont {Ehrichs}\ \emph {et~al.}(1995)\citenamefont
  {Ehrichs}, \citenamefont {Jaeger}, \citenamefont {Karczmar}, \citenamefont
  {Knight}, \citenamefont {Kuperman},\ and\ \citenamefont
  {Nagel}}]{Ehrichs1995}%
  \BibitemOpen
  \bibfield  {author} {\bibinfo {author} {\bibfnamefont {E.~E.}\ \bibnamefont
  {Ehrichs}}, \bibinfo {author} {\bibfnamefont {H.~M.}\ \bibnamefont {Jaeger}},
  \bibinfo {author} {\bibfnamefont {G.~S.}\ \bibnamefont {Karczmar}}, \bibinfo
  {author} {\bibfnamefont {J.~B.}\ \bibnamefont {Knight}}, \bibinfo {author}
  {\bibfnamefont {V.~Y.}\ \bibnamefont {Kuperman}}, \ and\ \bibinfo {author}
  {\bibfnamefont {S.~R.}\ \bibnamefont {Nagel}},\ }\href@noop {} {\bibfield
  {journal} {\bibinfo  {journal} {Science}\ }\textbf {\bibinfo {volume}
  {267}},\ \bibinfo {pages} {1632} (\bibinfo {year} {1995})}\BibitemShut
  {NoStop}%
\bibitem [{\citenamefont {Holmes}(1982)}]{Holmes1982}%
  \BibitemOpen
  \bibfield  {author} {\bibinfo {author} {\bibfnamefont {P.~J.}\ \bibnamefont
  {Holmes}},\ }\href@noop {} {\bibfield  {journal} {\bibinfo  {journal} {J.
  Sound Vib.}\ }\textbf {\bibinfo {volume} {84}},\ \bibinfo {pages} {173}
  (\bibinfo {year} {1982})}\BibitemShut {NoStop}%
\bibitem [{\citenamefont {Luck}\ and\ \citenamefont {Mehta}(1993)}]{Luck1993}%
  \BibitemOpen
  \bibfield  {author} {\bibinfo {author} {\bibfnamefont {J.~M.}\ \bibnamefont
  {Luck}}\ and\ \bibinfo {author} {\bibfnamefont {A.}~\bibnamefont {Mehta}},\
  }\href@noop {} {\bibfield  {journal} {\bibinfo  {journal} {Phys. Rev. E}\
  }\textbf {\bibinfo {volume} {48}},\ \bibinfo {pages} {3988} (\bibinfo {year}
  {1993})}\BibitemShut {NoStop}%
\bibitem [{\citenamefont {Luo}\ and\ \citenamefont {Han}(1996)}]{Luo1996}%
  \BibitemOpen
  \bibfield  {author} {\bibinfo {author} {\bibfnamefont {A.~C.~J.}\
  \bibnamefont {Luo}}\ and\ \bibinfo {author} {\bibfnamefont {R.~P.~S.}\
  \bibnamefont {Han}},\ }\href@noop {} {\bibfield  {journal} {\bibinfo
  {journal} {Nonlinear Dyn.}\ }\textbf {\bibinfo {volume} {10}},\ \bibinfo
  {pages} {1} (\bibinfo {year} {1996})}\BibitemShut {NoStop}%
\bibitem [{\citenamefont {Vogel}\ and\ \citenamefont {Linz}(2011)}]{Vogel2011}%
  \BibitemOpen
  \bibfield  {author} {\bibinfo {author} {\bibfnamefont {S.}~\bibnamefont
  {Vogel}}\ and\ \bibinfo {author} {\bibfnamefont {S.~J.}\ \bibnamefont
  {Linz}},\ }\href@noop {} {\bibfield  {journal} {\bibinfo  {journal} {Int. J.
  Bifurcat. Chaos}\ }\textbf {\bibinfo {volume} {21}},\ \bibinfo {pages} {869}
  (\bibinfo {year} {2011})}\BibitemShut {NoStop}%
\bibitem [{\citenamefont {Dorbolo}\ \emph {et~al.}(2005)\citenamefont
  {Dorbolo}, \citenamefont {Volfson}, \citenamefont {Tsimring},\ and\
  \citenamefont {Kudrolli}}]{Dorbolo2005}%
  \BibitemOpen
  \bibfield  {author} {\bibinfo {author} {\bibfnamefont {S.}~\bibnamefont
  {Dorbolo}}, \bibinfo {author} {\bibfnamefont {D.}~\bibnamefont {Volfson}},
  \bibinfo {author} {\bibfnamefont {L.}~\bibnamefont {Tsimring}}, \ and\
  \bibinfo {author} {\bibfnamefont {A.}~\bibnamefont {Kudrolli}},\ }\href@noop
  {} {\bibfield  {journal} {\bibinfo  {journal} {Phys. Rev. Lett.}\ }\textbf
  {\bibinfo {volume} {95}},\ \bibinfo {pages} {044101} (\bibinfo {year}
  {2005})}\BibitemShut {NoStop}%
\bibitem [{\citenamefont {Dorbolo}\ \emph {et~al.}(2009)\citenamefont
  {Dorbolo}, \citenamefont {Ludewig},\ and\ \citenamefont
  {Vandewalle}}]{Dorbolo2009}%
  \BibitemOpen
  \bibfield  {author} {\bibinfo {author} {\bibfnamefont {S.}~\bibnamefont
  {Dorbolo}}, \bibinfo {author} {\bibfnamefont {F.}~\bibnamefont {Ludewig}}, \
  and\ \bibinfo {author} {\bibfnamefont {N.}~\bibnamefont {Vandewalle}},\
  }\href@noop {} {\bibfield  {journal} {\bibinfo  {journal} {New J. Phys.}\
  }\textbf {\bibinfo {volume} {11}},\ \bibinfo {pages} {033016} (\bibinfo
  {year} {2009})}\BibitemShut {NoStop}%
\bibitem [{\citenamefont {Kubo}\ \emph {et~al.}(2015)\citenamefont {Kubo},
  \citenamefont {Inagaki}, \citenamefont {Ichikawa},\ and\ \citenamefont
  {Yoshikawa}}]{Kubo2015}%
  \BibitemOpen
  \bibfield  {author} {\bibinfo {author} {\bibfnamefont {Y.}~\bibnamefont
  {Kubo}}, \bibinfo {author} {\bibfnamefont {S.}~\bibnamefont {Inagaki}},
  \bibinfo {author} {\bibfnamefont {M.}~\bibnamefont {Ichikawa}}, \ and\
  \bibinfo {author} {\bibfnamefont {K.}~\bibnamefont {Yoshikawa}},\ }\href@noop
  {} {\bibfield  {journal} {\bibinfo  {journal} {Phys. Rev. E}\ }\textbf
  {\bibinfo {volume} {91}},\ \bibinfo {pages} {052905} (\bibinfo {year}
  {2015})}\BibitemShut {NoStop}%
\bibitem [{\citenamefont {McBennett}\ and\ \citenamefont
  {Harris}(2016)}]{McBennett2016}%
  \BibitemOpen
  \bibfield  {author} {\bibinfo {author} {\bibfnamefont {B.~G.}\ \bibnamefont
  {McBennett}}\ and\ \bibinfo {author} {\bibfnamefont {D.~M.}\ \bibnamefont
  {Harris}},\ }\href@noop {} {\bibfield  {journal} {\bibinfo  {journal}
  {Chaos}\ }\textbf {\bibinfo {volume} {26}},\ \bibinfo {pages} {093105}
  (\bibinfo {year} {2016})}\BibitemShut {NoStop}%
\bibitem [{\citenamefont {Der{\'e}nyi}\ \emph {et~al.}(1998)\citenamefont
  {Der{\'e}nyi}, \citenamefont {Tegzes},\ and\ \citenamefont
  {Vicsek}}]{Derenyi1998}%
  \BibitemOpen
  \bibfield  {author} {\bibinfo {author} {\bibfnamefont {I.}~\bibnamefont
  {Der{\'e}nyi}}, \bibinfo {author} {\bibfnamefont {P.}~\bibnamefont {Tegzes}},
  \ and\ \bibinfo {author} {\bibfnamefont {T.}~\bibnamefont {Vicsek}},\
  }\href@noop {} {\bibfield  {journal} {\bibinfo  {journal} {Chaos}\ }\textbf
  {\bibinfo {volume} {8}},\ \bibinfo {pages} {657} (\bibinfo {year}
  {1998})}\BibitemShut {NoStop}%
\bibitem [{\citenamefont {Farkas}\ \emph {et~al.}(1999)\citenamefont {Farkas},
  \citenamefont {Tegzes}, \citenamefont {Vukics},\ and\ \citenamefont
  {Vicsek}}]{Farkas1999}%
  \BibitemOpen
  \bibfield  {author} {\bibinfo {author} {\bibfnamefont {Z.}~\bibnamefont
  {Farkas}}, \bibinfo {author} {\bibfnamefont {P.}~\bibnamefont {Tegzes}},
  \bibinfo {author} {\bibfnamefont {A.}~\bibnamefont {Vukics}}, \ and\ \bibinfo
  {author} {\bibfnamefont {T.}~\bibnamefont {Vicsek}},\ }\href@noop {}
  {\bibfield  {journal} {\bibinfo  {journal} {Phys. Rev. E}\ }\textbf {\bibinfo
  {volume} {60}},\ \bibinfo {pages} {7022} (\bibinfo {year}
  {1999})}\BibitemShut {NoStop}%
\bibitem [{\citenamefont {Levanon}\ and\ \citenamefont
  {Rapaport}(2001)}]{Levanon2001}%
  \BibitemOpen
  \bibfield  {author} {\bibinfo {author} {\bibfnamefont {M.}~\bibnamefont
  {Levanon}}\ and\ \bibinfo {author} {\bibfnamefont {D.~C.}\ \bibnamefont
  {Rapaport}},\ }\href@noop {} {\bibfield  {journal} {\bibinfo  {journal}
  {Phys. Rev. E}\ }\textbf {\bibinfo {volume} {64}},\ \bibinfo {pages} {011304}
  (\bibinfo {year} {2001})}\BibitemShut {NoStop}%
\bibitem [{\citenamefont {Cai}\ and\ \citenamefont {Miao}(2019)}]{Cai2019}%
  \BibitemOpen
  \bibfield  {author} {\bibinfo {author} {\bibfnamefont {H.}~\bibnamefont
  {Cai}}\ and\ \bibinfo {author} {\bibfnamefont {G.}~\bibnamefont {Miao}},\
  }\href@noop {} {\bibfield  {journal} {\bibinfo  {journal} {Particuology}\
  }\textbf {\bibinfo {volume} {46}},\ \bibinfo {pages} {93} (\bibinfo {year}
  {2019})}\BibitemShut {NoStop}%
\bibitem [{\citenamefont {Bae}\ \emph {et~al.}(2004)\citenamefont {Bae},
  \citenamefont {Morgado}, \citenamefont {Veerman},\ and\ \citenamefont
  {Vasconcelos}}]{Bae2004}%
  \BibitemOpen
  \bibfield  {author} {\bibinfo {author} {\bibfnamefont {A.~J.}\ \bibnamefont
  {Bae}}, \bibinfo {author} {\bibfnamefont {W.~A.~M.}\ \bibnamefont {Morgado}},
  \bibinfo {author} {\bibfnamefont {J.}~\bibnamefont {Veerman}}, \ and\
  \bibinfo {author} {\bibfnamefont {G.~L.}\ \bibnamefont {Vasconcelos}},\
  }\href@noop {} {\bibfield  {journal} {\bibinfo  {journal} {Physica A}\
  }\textbf {\bibinfo {volume} {342}},\ \bibinfo {pages} {22} (\bibinfo {year}
  {2004})}\BibitemShut {NoStop}%
\bibitem [{\citenamefont {Halev}\ and\ \citenamefont
  {Harris}(2018)}]{Halev2018}%
  \BibitemOpen
  \bibfield  {author} {\bibinfo {author} {\bibfnamefont {A.}~\bibnamefont
  {Halev}}\ and\ \bibinfo {author} {\bibfnamefont {D.~M.}\ \bibnamefont
  {Harris}},\ }\href@noop {} {\bibfield  {journal} {\bibinfo  {journal}
  {Chaos}\ }\textbf {\bibinfo {volume} {28}},\ \bibinfo {pages} {096103}
  (\bibinfo {year} {2018})}\BibitemShut {NoStop}%
\bibitem [{\citenamefont {Astumian}\ and\ \citenamefont
  {Bier}(1994)}]{Astumian1994}%
  \BibitemOpen
  \bibfield  {author} {\bibinfo {author} {\bibfnamefont {R.~D.}\ \bibnamefont
  {Astumian}}\ and\ \bibinfo {author} {\bibfnamefont {M.}~\bibnamefont
  {Bier}},\ }\href@noop {} {\bibfield  {journal} {\bibinfo  {journal} {Phys.
  Rev. Lett.}\ }\textbf {\bibinfo {volume} {72}},\ \bibinfo {pages} {1766}
  (\bibinfo {year} {1994})}\BibitemShut {NoStop}%
\end{thebibliography}%

\end{document}